\documentstyle[preprint,aps,prl]{revtex}
\tightenlines 
\begin{document} 

\title{Short-time critical dynamics of the Baxter-Wu model}

\author{Everaldo Arashiro \thanks{everaldo@dfm.ffclrp.usp.br},
	J.R. Drugowich de Fel\'{\i}cio \thanks{drugo@usp.br}
	} 

\address{Departamento de F\'{\i}sica e Matem\'atica\\
Faculdade de Filosofia, Ci\^encias e Letras de Ribeir\~ao Preto.\\
Universidade de S\~ao Paulo\\
Av. Bandeirantes, 3900 - CEP - 14040-901\\
Ribeir\~ao Preto - S\~ao Paulo - Brazil}
\maketitle

\begin{abstract}
We study the early time behavior of the Baxter-Wu model, an Ising model with three-spin 
interactions on a triangular lattice. Our estimates for the dynamic
exponent $z$ are compatible with results recently obtained for two models which belong 
to the same universality class of the Baxter-Wu model: the two-dimensional four-state 
Potts model and the Ising model with three-spin interactions in one direction. However, 
our estimates for the dynamic exponent $\theta $ of the Baxter-Wu model are completely 
different from the values obtained for those models. This discrepancy could be related 
to the absence of a marginal operator in the Baxter-Wu model.


\end{abstract}
\newpage

\section{Introduction}

Since the work of Janssen {\it et al.} \cite{Janssen} and Huse \cite{Huse} pointing out 
that universality and scaling are present at the early stage of the time evolution
of dynamic systems, several statistical models have been investigated using
Monte Carlo simulations in the short-time regime \cite{Li}. This kind of investigation 
avoids the critical slowing down effects and provides an alternate method for calculating the 
dynamic exponent $z$, the static exponents $\beta$ and $\nu$ as well as the new dynamic 
exponent $\theta$ that governs the anomalous behavior of the nonequilibrium magnetization 
when the system is quenched to the critical temperature $T_c$.
Using this new approach kinetic Ising and Potts models 
\cite{Okano,Claudia1,Ye,Roberto1,Zhengrev}
were exaustively studied in order to confirm available results and to enlarge our knowledge 
of critical phenomena, including the critical relaxation of the early time dynamics.
 
In this paper, we revisit the Ising model with three-spin interactions on a
triangular lattice, called Baxter-Wu (BW) model
after its exact solution obtained thirty years ago \cite{BW}. The
motivation arises from the disagreement between results for the exponent $z$ obtained for the BW
model and for two other models - the two-dimensional four-state Potts model and
the Ising model with three-spin interactions in one direction -
which are known to belong to the same universality class.
The BW model consists of Ising spins ($S_{i}=\pm 1$) on a
triangular lattice with the Hamiltonian 

\begin{equation}
H=-J\sum_{<i,j,k>}S_{i}S_{j}S_{k} ,
\end{equation}
where the sum involves the product of the spins on each triangle of the lattice.
As the Ising model on a square lattice this model undergoes a phase transition at 
the critical temperature $%
K_{c}=J/kT_{c}=0.5\ln (1+\sqrt{2)}$. The ground-state is fourfold degenerate (see Fig. 1) and
the relevant symmetry is semi-global (the Hamiltonian is invariant under
reversal of all of the spins belonging to two of three sublatices into
which the original lattice can be decomposed). The BW model exhibits the same
static critical exponents ($\alpha =\nu =2/3$ and $\eta=1/4$) \cite{Wu} as those of the 
two-dimensional four-state Potts model and of the Ising model with three-spin
interactions (IMTSI) in one direction \cite{Turban}. However, finite-size studies of 
the BW model have good convergence, without running into the known difficulties encountered 
in the studies of these other models. Several authors \cite{Alcaraz} have attributed that difference to
a dangerous marginal operator present in the 4-state Potts and IMTSI models
which does not have a counterpart in the BW model.

Although the static critical behavior of these systems is well understood,
only a few results are known regarding dynamical properties. For
instance, much work it is needed to unravel the interesting question of a
possible extended universality, {\it i. e.}, if these three models (BW, four-state
Potts and the IMTSI) exhibit the same exponent $z$. Previous estimates of
the critical exponent $z$ for the four-state Potts model
are scattered between 2 and 4 \cite{Jan}. The most recent results however suggest
values close to 2.3 \cite{Roberto2,Claudia2}. Short-time 
simulations for the BW model were first performed by Santos and Figueiredo 
\cite{Figueiredo}, who found $z=2.07(1)$, a very small value when compared to the values 
obtained for the same exponents of the four-state Potts and IMTSI models 
\cite{Roberto2,Claudia2}. Moreover, they stress that this value of $z$ is very close to the 
dynamic exponent of the the two-dimensional (2D) Ising model when Glauber dynamics is used. For
the new dynamic exponent $\theta $ related to the critical initial slip of the order parameter,
 situation is even worse. Okano {\it et al} have conjectured \cite{Okano} that for
the four-state Potts model the exponent \ $\theta $ should be negative and close to zero.
 This result was only recently confirmed \cite{Roberto3}. For the IMTSI
model, two independent estimates were obtained \cite{Claudia2,Zhang} 
and fall in the range $ -0.04< \theta < 0.07$ , whereas estimates for the
BW model are not available.

Motivated by our recent results for the IMTSI model \cite{Claudia2}, we
decided to investigate the short-time critical dynamics of the BW
model in order to check the value of the dynamic exponent $z$ and the
utility of the exponent $\theta $ in detecting the presence of a marginal
operator \cite{Claudia1,Ye,Oerding,Claudia e Tania}. In the next section we
calculate the exponent $\theta $ studying the evolution of the order parameter
 for different values of the initial sublattice magnetization $ m_0 = (3/N)\sum_i S_i (0)$ 
where the index $i$ runs over only one kind of sites of the triangular lattice (circle, 
square or star points in Fig. 1) and $N$ is the number of spins. In addition, 
we estimate the same exponent calculating the time correlation of the order 
parameter \cite{Tania}. In section III we summarize our results for the dynamic critical 
exponent $z$. In section IV we list estimates 
for the correlation length exponent $\nu $ and for the index $\beta $ of the 
magnetization. Finally, in section V we present our conclusions.

\section{The dynamic exponent $\protect\theta $}

Janssen {\it et al.}\cite{Janssen} have shown on the basis of the renormalization group 
theory that, at an early stage of the time evolution, quantities such as the magnetization and 
its moments follow a power law when the parameters of a dynamical system are adjusted
to their critical values. The exponent \ $\theta $ which characterizes the time evolution of the 
magnetization is independent of the known set of
static critical exponents and even from the dynamic exponent $z$. However, like other exponents
 $\theta$ seems to depend on general properties such as
dimensionality of the spin and either symmetry of the Hamiltonian or the 
evolution rules. Such universal behavior was found in two and
three-dimensional Ising models \cite{Okano,Barreto}, two-dimensional three
and four-state Potts models \cite{Roberto2,Potts3} as well as 
in irreversible models with synchronous and continuous dynamics 
 \cite{Tania,Mendes,Tania e Drugo}. In addition, nonuniversal behavior 
was detected for the exponent $\theta $ \cite{Claudia1,Ye,Ashkin-Teller,Claudia e
Tania} although always accompanied by a marginal operator.

In this paper we estimate the dynamic exponent \ $\theta $ by two different
techniques. First we quench the system from a high temperature
to $T_{c}$ and follow the nonequilibrium relaxation of various realizations of the process 
with the same initial sublattice magnetization $m_0$. In Fig. 2 we show the behavior of 
the order parameter when $m_{0}=0.01$. As proposed by Janssen {\it et al} \cite{Janssen} 
it follows the power law

\begin{equation}
M(t)\simeq m_{0}t^{\theta },  \label{m0}
\end{equation}%
but here $\theta$ is negative. Figure 3 exhibits the behavior of the
exponent $\theta $ for different values of $m_{0}$, as well as a linear fit
which leads to the value $\theta=-0.186 \pm 0.002$.

\ The second technique consists of studying the time correlation 
of the magnetization

\begin{equation}
C(t)=\left\langle \sum_{i}S_{i}(t)\sum_{j}S_{j}(0)\right\rangle ,
\label{tania}
\end{equation}%
which was shown \cite{Tania} to behave as $t^{\theta }$ when $<M(0)> =0$. In Fig. 4 
we show the time dependence of the time correlation C. We
stress that this approach was shown to be valid when the evolution matrix
has up-down symmetry which is not valid in the present case. However, it is possible to
show \cite{Taniapvc} that the $ Z(2)\bigotimes Z(2)$ symmetry of the BW
Hamiltonian is sufficient to preserve the above mentioned result. The value 
of $\theta$ obtained with this technique ($\theta=-0.185 \pm 0.001$) corroborates the 
 value obtained directly from the behavior of the magnetization. We stress 
that the time correlation method \cite{Tania} is numerically more convenient since we  
do not need to work with specially prepared initial configurations (with fixed $m_{0}$) nor it is 
necessary to take the delicate limit 
$m_{0}\rightarrow 0$.

The results obtained with the two techniques are essentialy the same and fall in 
the range $-0.188\leq \theta \leq -0.184$. They are completely 
different from the value $\theta =-0.03 \pm 0.01$ encountered for the IMTSI model \cite{Claudia2} and 
from $\theta = -0.047 \pm 0.0033$ \cite{Roberto3} obtained recently for the four-state 
Potts model. Thus, based on our estimates for the exponent\ $\theta $ of the BW model, 
we conclude that its dynamic behavior at short-time is completely different from 
that of the IMTSI and four-state Potts models albeit both models have the same static 
exponents as the BW model. Thus, at least in this case, the 
exponent $\theta $ seems to be able to capture the subtle difference between models 
that have the same leading exponents but not the same corrections to scaling.

\section{The dynamic critical exponent $z$}

\bigskip Santos and Figueiredo \cite{Figueiredo} obtained the
exponent $z$ of the BW model by means of the second
order cumulant $U_{2}$ \cite{Zhengrev} 

\begin{equation}
U_{2}= \frac{<M^{2}>}{<M>^{2}} -1 , \label{Zheng}
\end{equation}%
starting from completely ordered samples ($m_{0}=1$). As claimed by Zheng, $U_{2}$ 
should behave at the critical temperature as a power law

\begin{equation}
U_{2}\propto t^{d/z} ,
\end{equation}%
which would allow us to estimate $z$ without using any other critical exponents.
Santos and Figueiredo found $z = 2.07(1)$ for $m_{0}=1$ (all 
sub-lattices initially with spins up) and $1.96(2)$ for $m_{0}=-1/3$, {\it i.e.}, spins
are initially up in one sub-lattice and down in the other two. These
estimates, obtained with Glauber dynamics \cite{Glauber}, are very different
from results recently obtained for the four-state Potts model \cite{Roberto2}
and for the Ising model with three spin interactions in one direction \cite%
{Claudia2} which both yield values close to 2.3. 

This kind of disagreement has already appeared in the literature when the second
cumulant was used to obtain estimates for the exponent $z$ for the $3-$state Potts model 
\cite{Zhengrev} and the IMTSI model \cite{Claudia2}. Thus, in order to check if the BW
model (which also belongs to the $4-$state Potts model universality class)
exhibits indeed a different value for the dynamic exponent $z$, we
have calculated this exponent using three other techniques. To begin with, we
estimated $z$ by collapsing the generalized fourth-order Binder cumulant

\begin{equation}
U_{4}(t,\tau ,L)=1-\frac{<M^{4}>}{3<M^{2}>^{2}}  \label{Binder cumulant}
\end{equation}%
which should only depend on the ratio $t/L^{z}$ when $T=T_{c}$, 
according to scaling laws valid in the beginning of the evolution. In Fig. 5
we show the Binder cumulant as a function of the time for two different
lattices $L=96$ and $L^{\prime }=192$, both starting from
ordered states ($m_{0}=1$). We also plot in Fig. 5 the cumulant for the
larger lattice against the re-scaled time $t^{\prime }=2^{z}t$ where $z$ is defined
via the relation

\begin{equation}
U_{4}(2^{z}t,0,L^{\prime })=U_{4}(t,0,L).  \label{invariant}
\end{equation}

After interpolating the re-scaled curve for the lattice $L^{\prime
}$ , the value of $z $ 
is selected using the $\chi ^{2}$ test \cite{Numerical Recipes} yielding the 
best estimate $2.285(10)$. We applied the same analysis to
the cumulant \cite{Zhengrev} 
\begin{equation}
{\tilde U} = \frac{<M^{2}>}{\left\vert M\right\vert ^{2}} -1,  \label{modulo}
\end{equation}%
using ($m_{0}=1$) as shown in Fig. 6. The
result in this case is $z=2.29(1)$.

Finally, we used mixed initial conditions in order to obtain the function $F_{2}(t)$, 
given by

\begin{equation}
F_{2}(t)=\frac{<M^{2}>_{m_{0}=0}}{<M>_{m_{0}=1}^{2}} , \label{mixed}
\end{equation}%
which was shown \cite{Roberto2} to behave as $t^{d/z}$. This approach was recently proposed 
in order to obtain the exponent $z$ with high precision. Actually, curves of $F_{2}$ are 
straight lines (see Fig. 7) which do not depend on the lattice size (one does not need to 
rescale the time). Our estimate for $z$ using Eq.(\ref{mixed}) and taking values at each 5 
steps is $\ z=2.294(6)$. This value is in complete agreement with the result $z=2.290(3)$ 
recently obtained \cite{Roberto2} for the four-state Potts model and corroborates
 the above results obtained by different procedures.

It is worth stressing that all the results shown until
now were obtained with heat-bath updating whereas Santos and Figueiredo \cite%
 {Figueiredo} performed simulations using Glauber dynamics. In order to
rule out any influence of the dynamics on the results we repeated our
simulations using Glauber updating and verified that our result remain essentially the
same. In addition, we have used heat-bath updating to calculate
the time dependence of the second order cumulant (Eq.(\ref{Zheng})) where
the average is always obtained from initially ordered samples ($m_{0}=1$).
Our best estimate for $z$ is $2.03\pm 0.01$, which agrees reasonably well
with the result by Santos and Figueiredo, indicating that this severe
disagreement stems from a difficulty from the scaling for the second
cumulant method.

\section{The thermal exponents $\protect\nu $ and $\protect\beta $}

Taking the logarithmic derivative of the magnetization \cite{Zhengrev}

\begin{equation}
M(t,\tau )=t^{-\beta /\nu z}M(1,t^{1/\nu z}\tau )
\end{equation}%
with respect to $\tau =(T-T_{c})/T_c$, we obtain

\begin{equation}
\partial _{\tau }\ln M(t,\tau )=t^{1/\nu z}\partial _{\tau ^{\prime }}\ln
(F(\tau ^{\prime })\mid _{\tau ^{\prime }=0} , \label{derivative}
\end{equation}%
which follows a power law that does not depend on $L$, where $F(\tau ^{\prime})$ is a
 scaling function. Thus, a log-log plot of that derivative {\it versus} time gives 
 $ 1/ \nu z $ (see Fig. 8). Once we have obtained 
the dynamical exponent $z$, the thermal exponent $\nu $ may be determined. Our estimate 
for the exponent $\nu $ in the short-time analysis is $0.65\pm 0.01$ which should 
be compared to the exact value $2/3$ from Ref. \cite{Baxter}. 

Next, we follow the decay of the sublattice magnetization in
initially ordered samples ($m_{0}=1$). The scaling law

\begin{equation}
M\propto t^{-\beta /\nu z}
\end{equation}%
allows us to obtain $\beta /\nu z$ which in turn leads to the exponent $%
\beta $ by using the previous result for the product $\nu z$. We show a
log-log plot of $M$ {\it versus} $t$ in Fig. 9. A linear fit gives the value $\
0.0527\pm 0.0005$ for the ratio $\beta /\nu z$ leading to the estimate 

\begin{equation}
\beta =0.079\pm 0.002 .
\end{equation}%
We stress that our result for $\beta $ is in fair agreement with the exact result  $1/12$ 
\cite{BW}.

Before closing this section we present another estimate for the exponents $%
\beta $ and $\nu $ obtained from  caling of the so-called Baxter order
parameter $M_{B}$ given by
\begin{equation}
M_{B}=\sum\limits_{<ijk>} S_{i}S_{j}S_{k} ,
\end{equation}%
where the sum is done over all triangles of the lattice and the indices $i$, $j$ 
and $k$ denote sites composing an elementary triangle (plaquette). Barber and Baxter \cite{Barber e Baxter} 
have shown that $M_{B}$ is zero for
temperatures below $T_{c}$ and different from zero above $T_{c}$. In this
sense $M_{B}$ functions as a disorder parameter \cite{Kadanoff} which goes to zero
when $T\rightarrow T_{c}$ from above. However, this new 
quantity has the same critical exponent 
as the sublattice magnetization. 
Thus, the exponent $\beta$ can be obtained following
the evolution of the Baxter's order parameter. The advantage is that we do not
need to divide the system into sub-lattices. In Fig. 10 we show the decay
of the Baxter order parameter as well as the value of the ratio $\beta /\nu z
$. In order to eliminate the product $\nu z$ from our result we start again
from ordered samples ($M_{B}=1$) and use Eq.(\ref{derivative}) replacing $M$
by $M_{B}$ to obtain $1/\nu z$. Figure 11 shows the evolution of the
derivative. Our new estimate for $1/\nu z$ is $0.651\pm 0.006$, yelding 
for $\nu $ the value $0.67\pm 0.01$ \ in very good agreement with the exact 
result $2/3$ \cite{BW}. The new estimate for $\beta $ is $%
0.080\pm 0.002$ which is in fair agreement with our previous result and
should be compared to the pertinent result $1/12$.

\section{Conclusions}

We revisited the Baxter-Wu model using short-time Monte Carlo simulations.
Our best estimate for the dynamic critical exponent is $z =2.294\pm 0.006$. It was 
obtained by a recently proposed technique \cite{Roberto2} that combines simulations 
performed with different initial conditions. This result is compatible with recently 
obtained estimates for $z$ in the case of the two-dimensional four-state Potts model 
\cite{Roberto2}
as well as for the two-dimensional Ising model with three-spin interactions along 
one direction \cite{Claudia2}. The result for $z$ was confirmed by two other techniques 
based on the collapse of cumulants. In addition, using scaling relations for the order 
parameter and its derivative with respect to the reduced temperature $\tau$, 
we were also able to estimate the static critical exponents $\beta $ and $\nu $. We have also 
obtained the exponents $\beta $ and $\nu $ using the Baxter$^\prime$s order parameter. The 
results for these exponents are in good agreement with the exact values \cite{BW}. However, 
our estimates for the new dynamic exponent $\theta $ of the Baxter-Wu model are completely 
different from the values found for that index when studying the two-dimensional Ising model 
with three-spin interactions in one direction and the four-state Potts model. In this sense, 
the exponent $\theta$ seems to be able to capture subtle difference between models 
which exhibit the same leading critical exponents but different corrections to scaling. 
\bigskip

\hspace{7cm} {\bf Acknowledgments}

We would like to thank T. Tome for discussions about her result concerning the validity 
of the correlation method for the BW model prior to its publication. We would also like to 
thank R. Dickman, N. Caticha, N. Alves, U.H.E. Hansmann and N. Alves Jr. for discussions and careful 
reading of the manuscript. This work was supported by the Brazilian agencies FAPESP, CAPES 
and CNPq.

\newpage
{\LARGE Figure Captions}
\vspace{2.0cm}

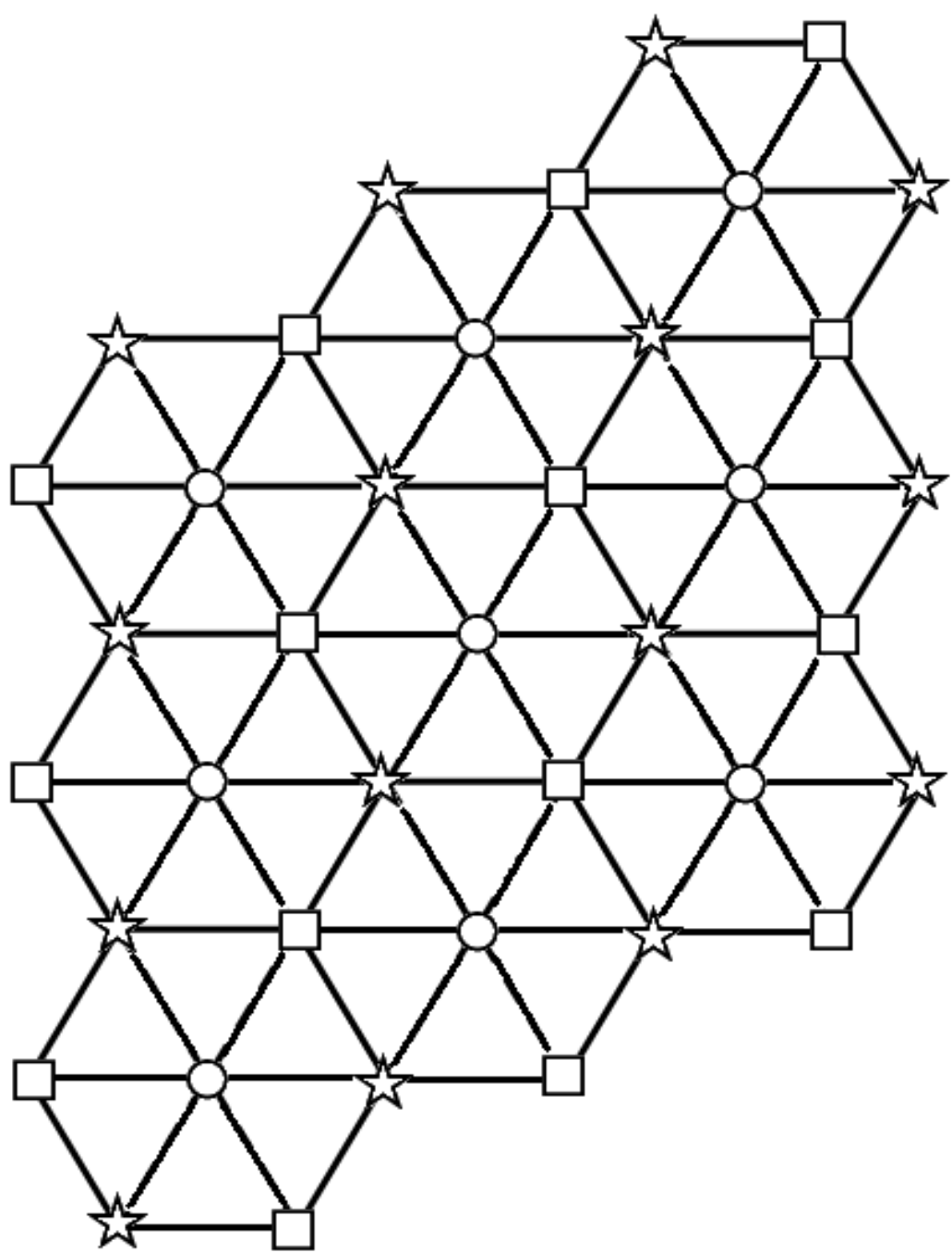
Figure 1. The Baxter-Wu model is defined on a triangular lattice where
 Ising spins ($S_i = \pm 1$ ) are coupled by nearest neighbor three-spin 
interactions. The lattice can be divided in three sub-lattices, characterized 
in the figure by circle, square and star points. Each symbol represents a class of
spins which does not interact with each other. The ground-state is
fourfold degenerate, corresponding to the three configurations with up
 spins in two sub-lattices and down spins in the other, besides the
obvious configuration of spins up in all the sublattices.

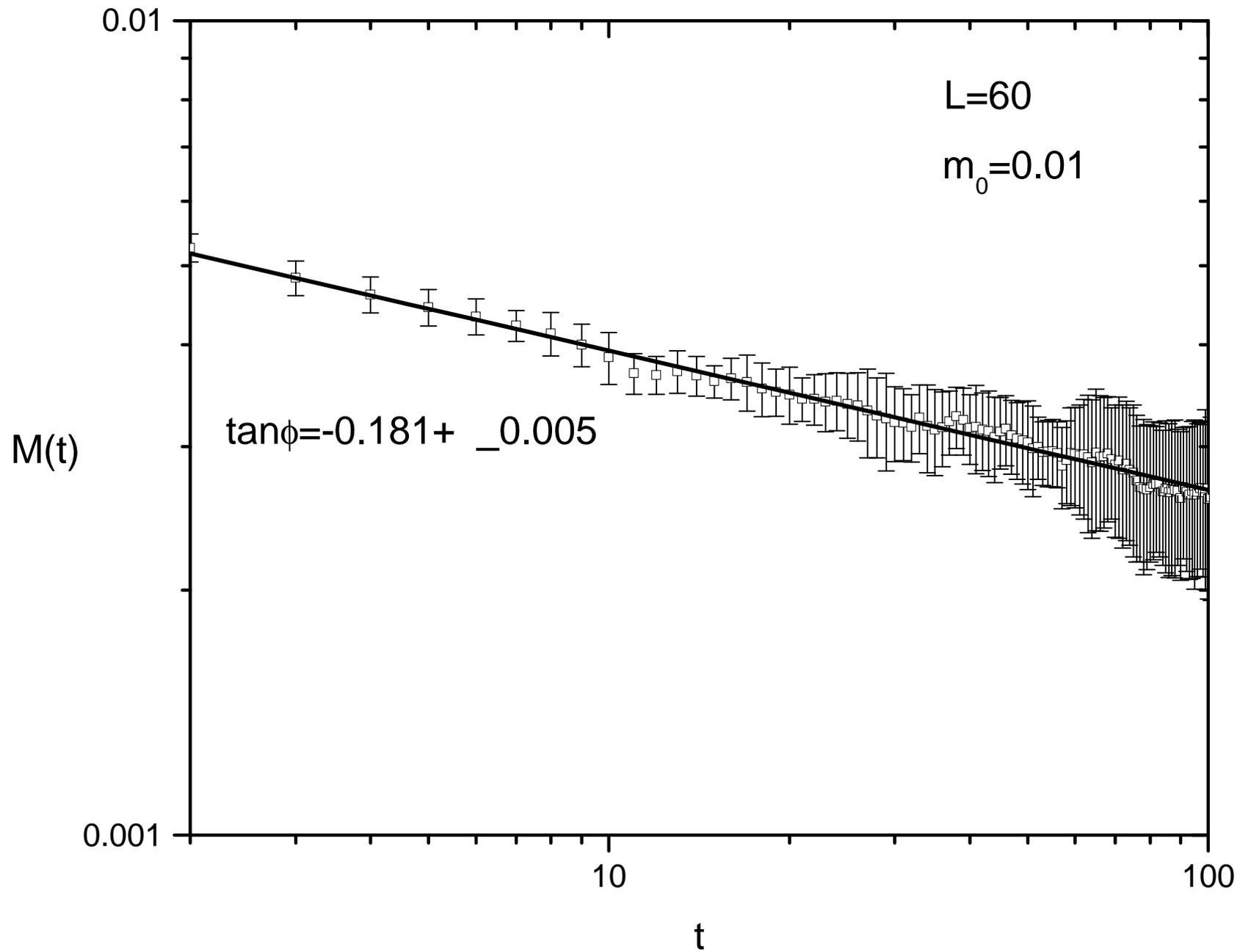
Figure 2. Power law behavior of the magnetization for samples with initial
magnetization $m_{0}=0.01$. The error bars were calculated over 10 sets of 60000
samples.

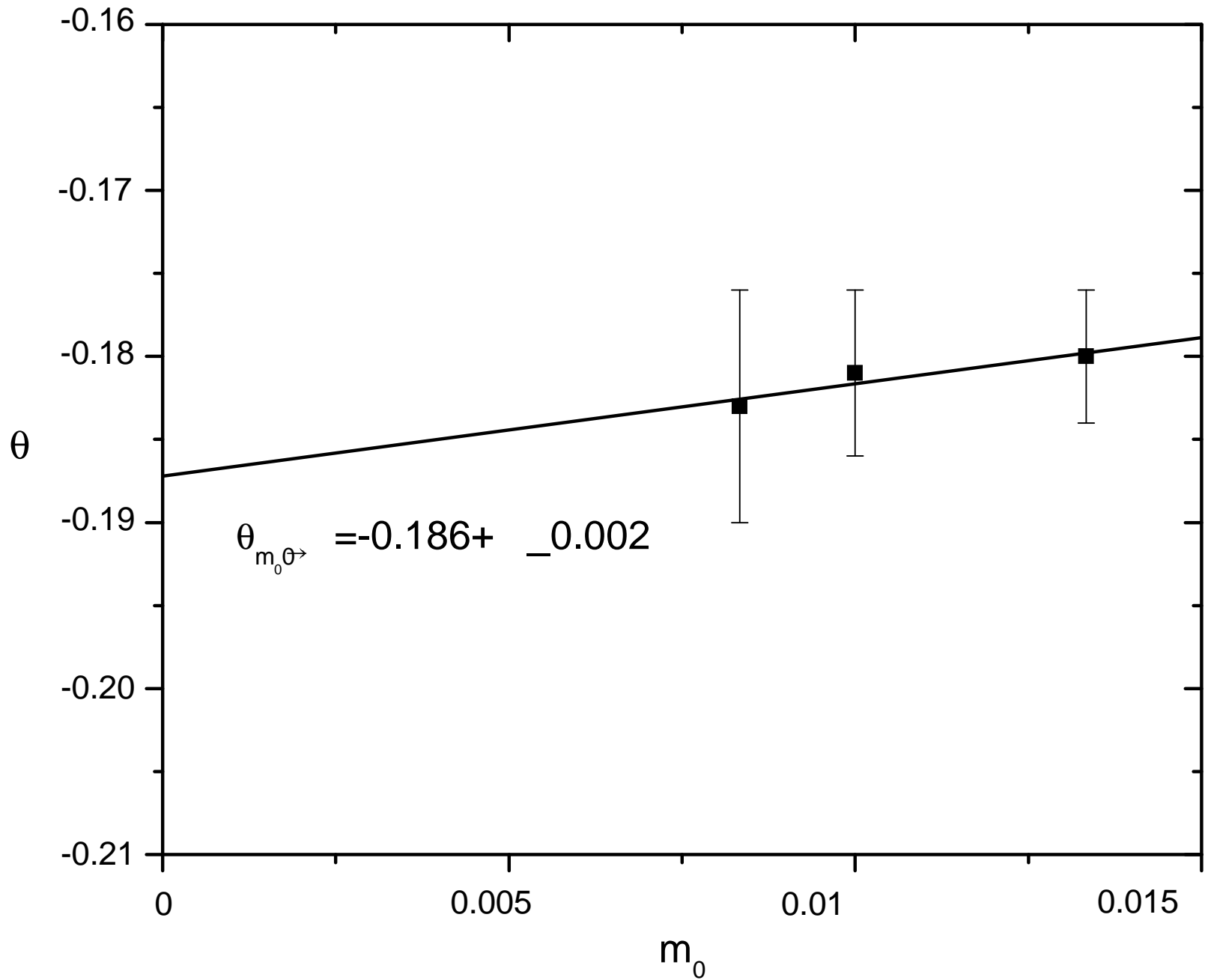
Figure 3. Dynamic exponent $\theta $ as a function of the initial
magnetization $m_{0}$ for the lattice size $L=60$. Each point 
represents an average over 10 sets of 60000 samples.

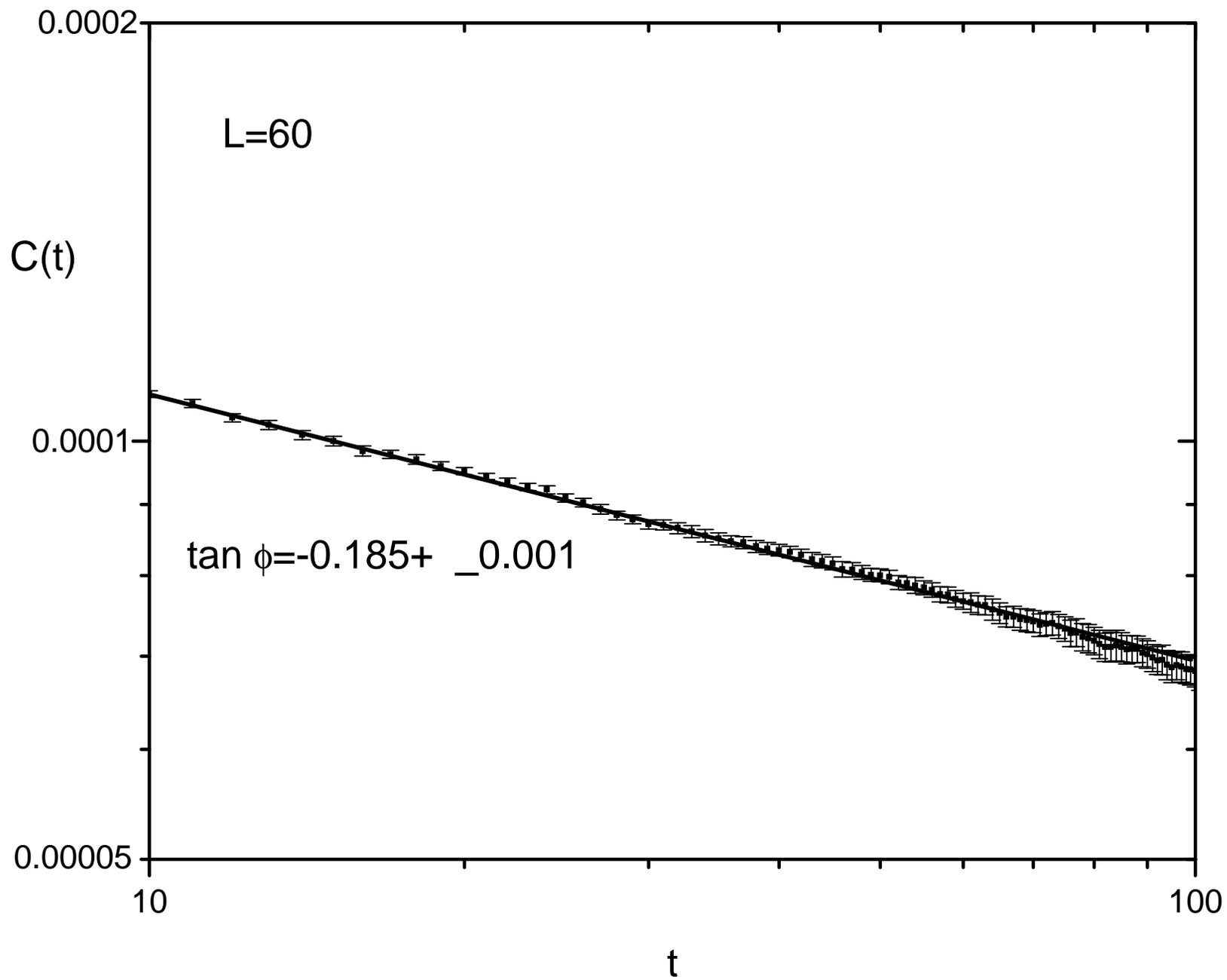
Figure 4. Time correlation of the total magnetization for samples with
 $<M(t=0)>=0$. Error bars were calculated over 10 sets of 60000 samples.

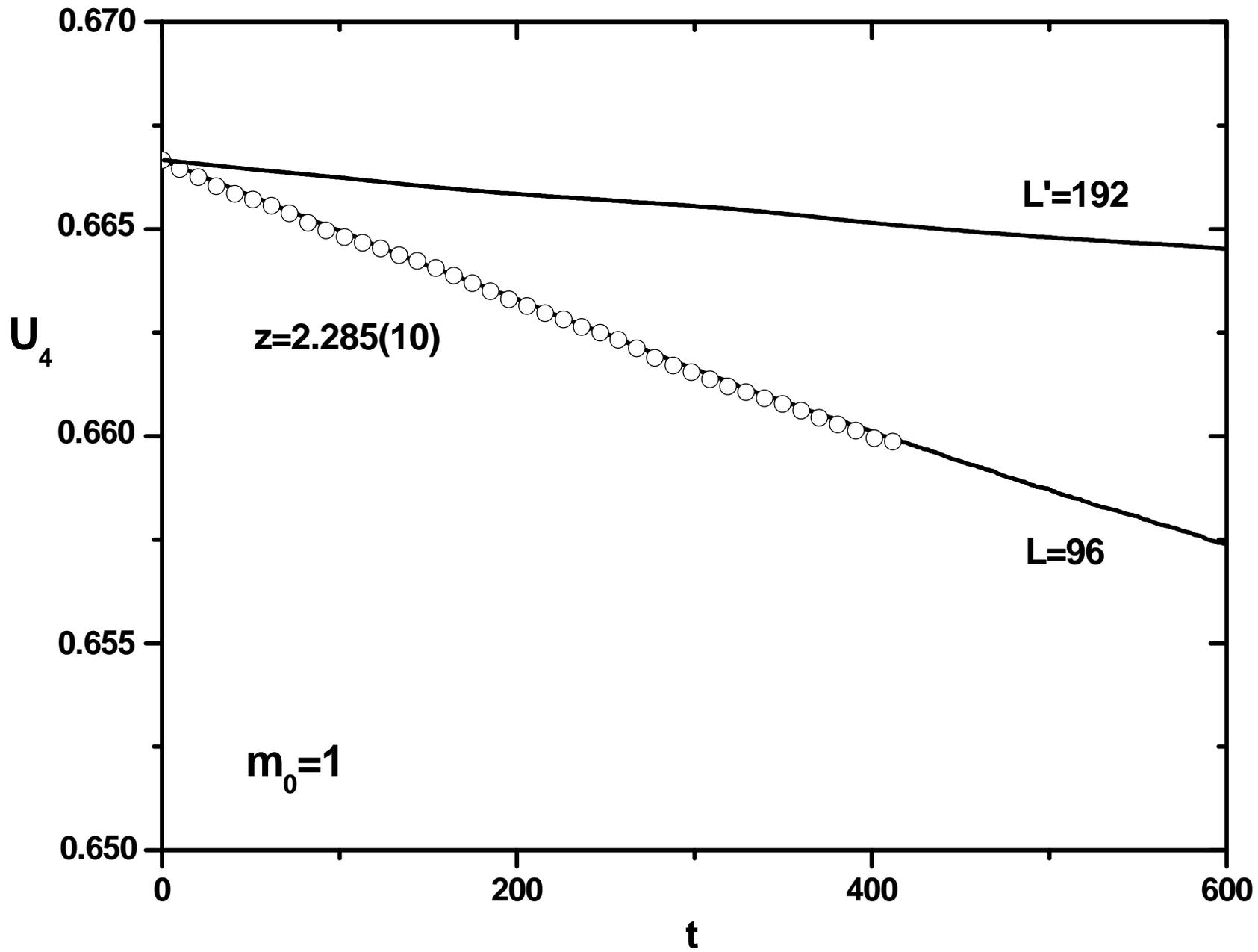
Figure 5. Binder cumulant $U_{4}(t,0,L)$ for lattice size $L=96$ and $%
L^{\prime }=192$ and initial magnetization $m_{0}=1$. The open circles on
the continuous line show the cumulant for the bigger lattice re-scaled in
time with $z$ obtained from Eq. (\ref{invariant}).

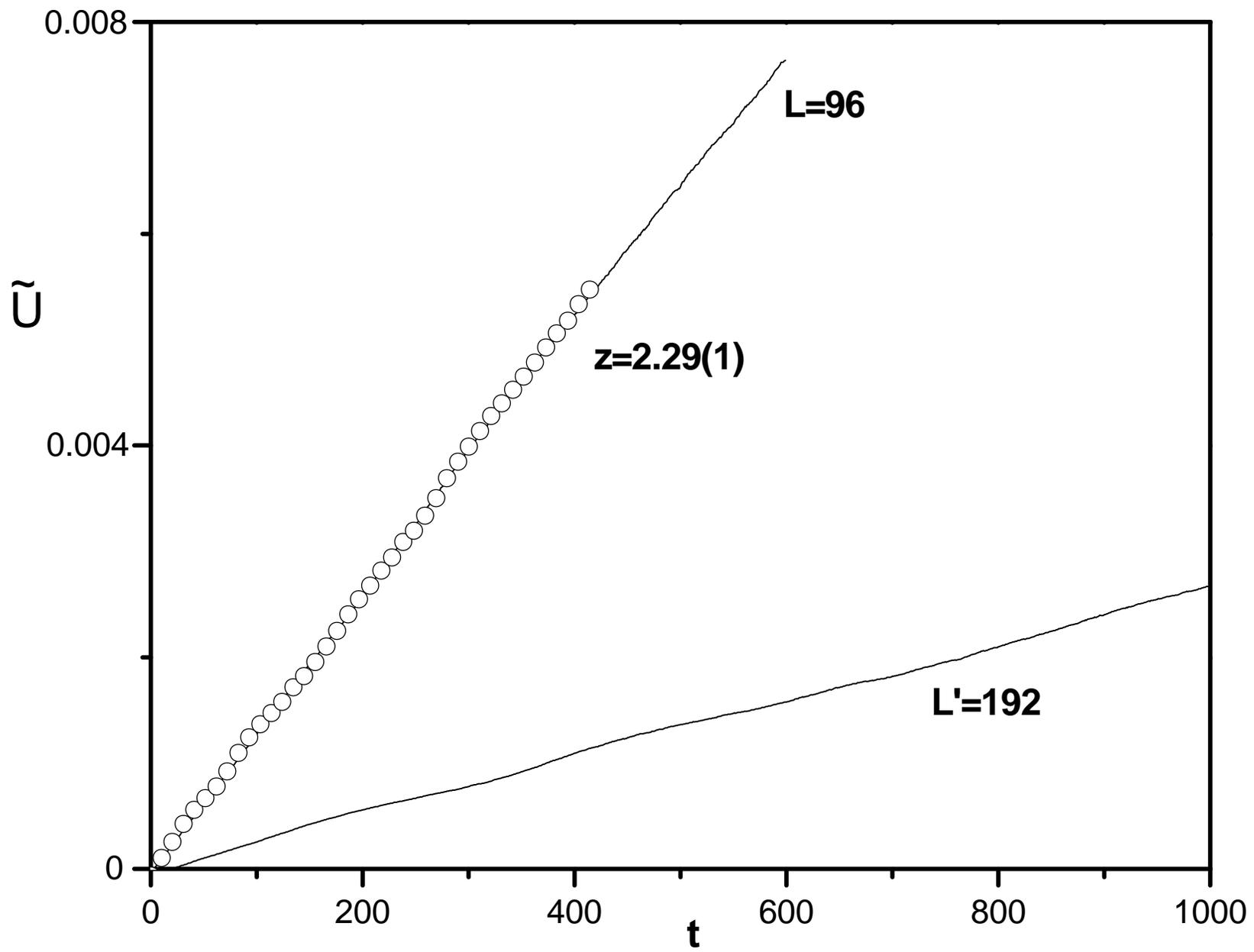
Figure 6. $\tilde{U}$ cumulant for $L=96$, $L^{\prime
}=192$ and initial magnetization $m_{0}=1$. The open circles correspond to
the re-scaled cumulant for the larger lattice with $z=2.29(1)$.

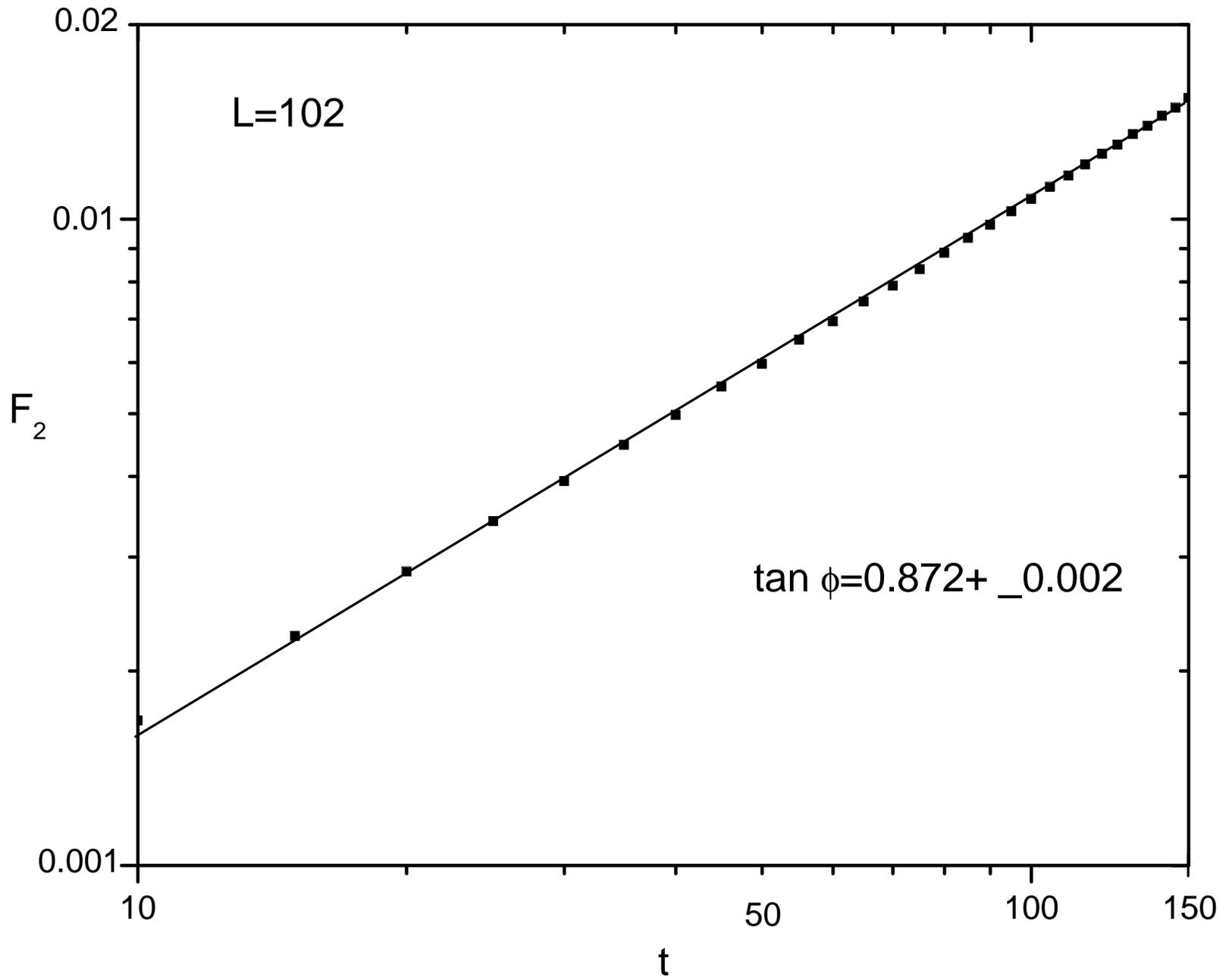
Figure 7. Time evolution of $F_{2}$ for $L=102$ with mixed initial
conditions [Eq.(\ref{mixed})]. The error bars, calculated over 10 sets of
30000 samples, are smaller than the size of the points.

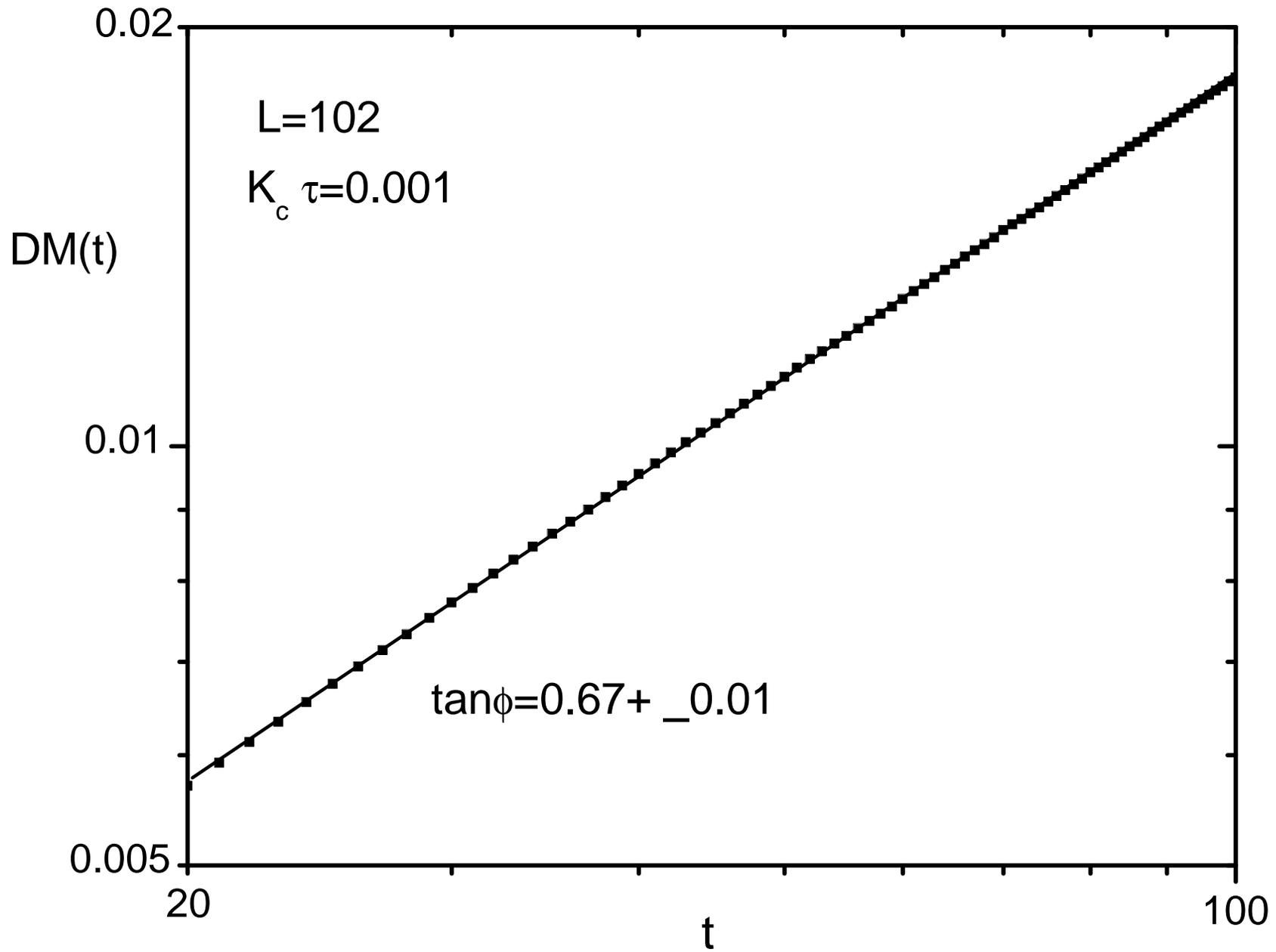
Figure 8. Logarithmic derivative of the magnetization with respect to the 
reduced temperature $DM(t)$ as a function of time. Data were obtained for 
the lattice size $L=102$ using temperature variation $K_c \tau =0.001$. 
The slope corresponds to the ratio $ 1/ \nu z$, after Eq. (\ref{derivative}). 
The error bars, calculated over 10 sets of 30000 samples, are smaller than 
the size of the points.

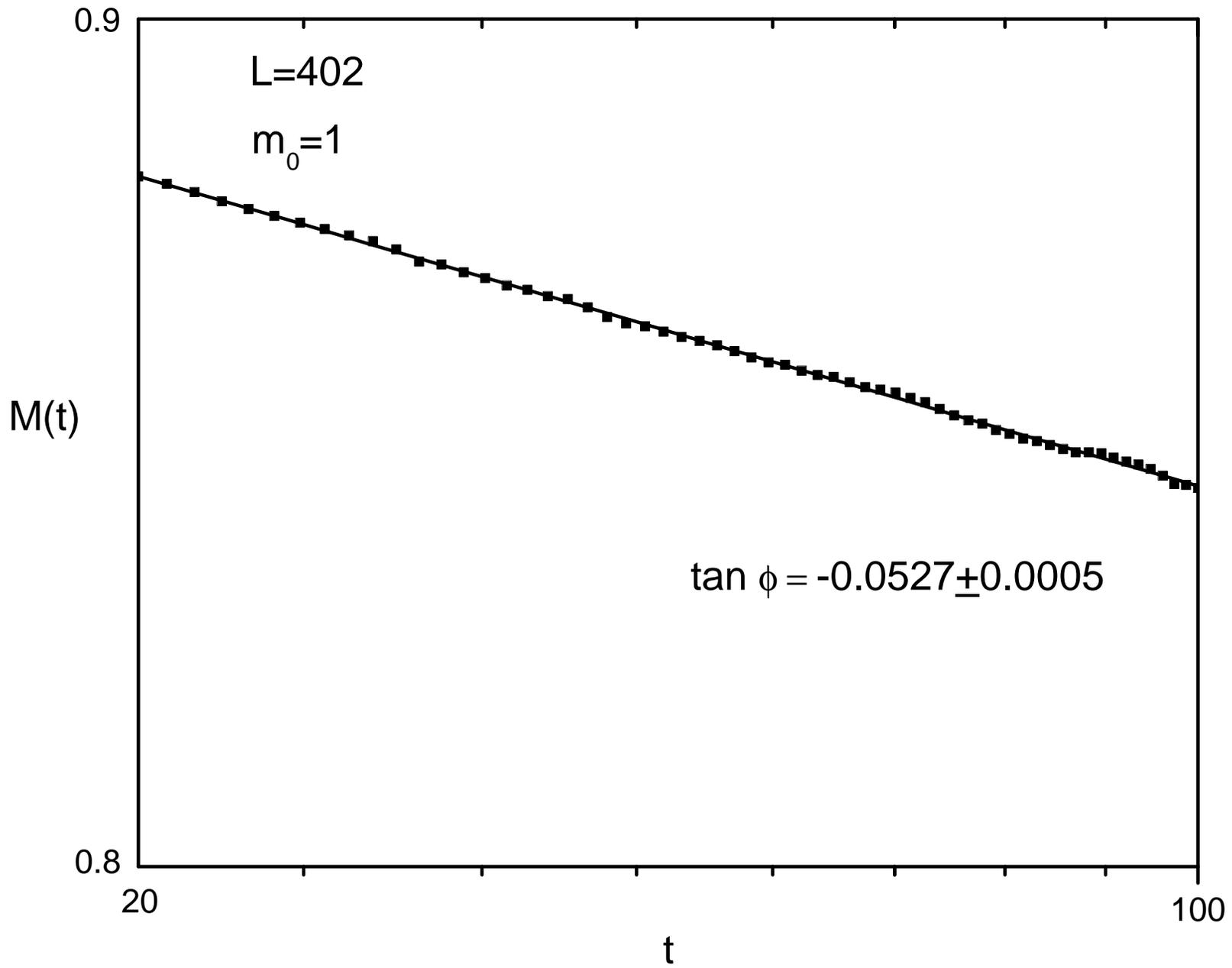
Figure 9. Log-log plot of the time evolution of the magnetization for
initially ordered samples ($m_{0}=1$). The error bars (calculated over 5 sets of
1000 samples) are smaller than the size of the points.

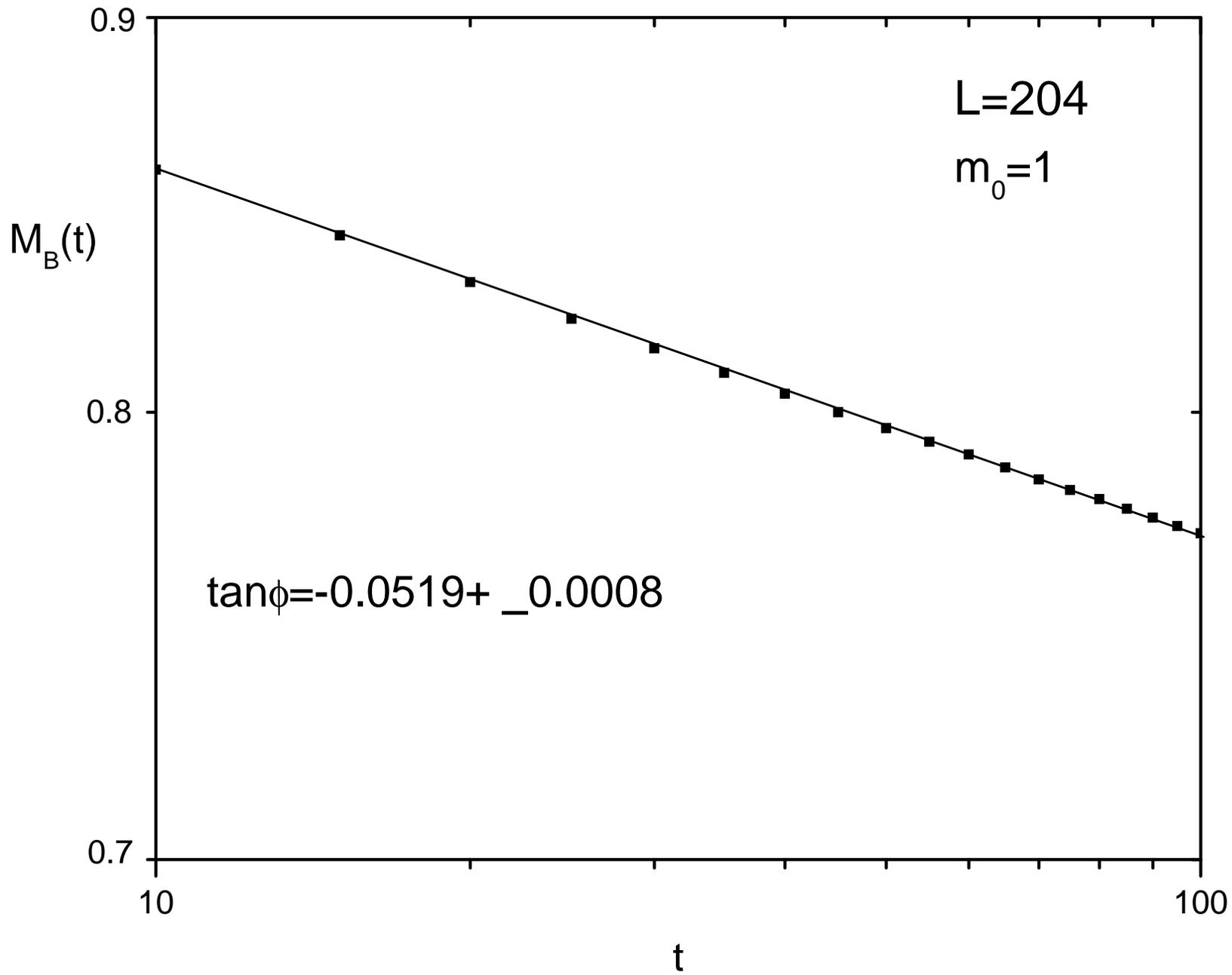
Figure 10. Time decay of the Baxter's order parameter $M_{B}$. Error bars
were calculated over 5 sets of 10000 samples.

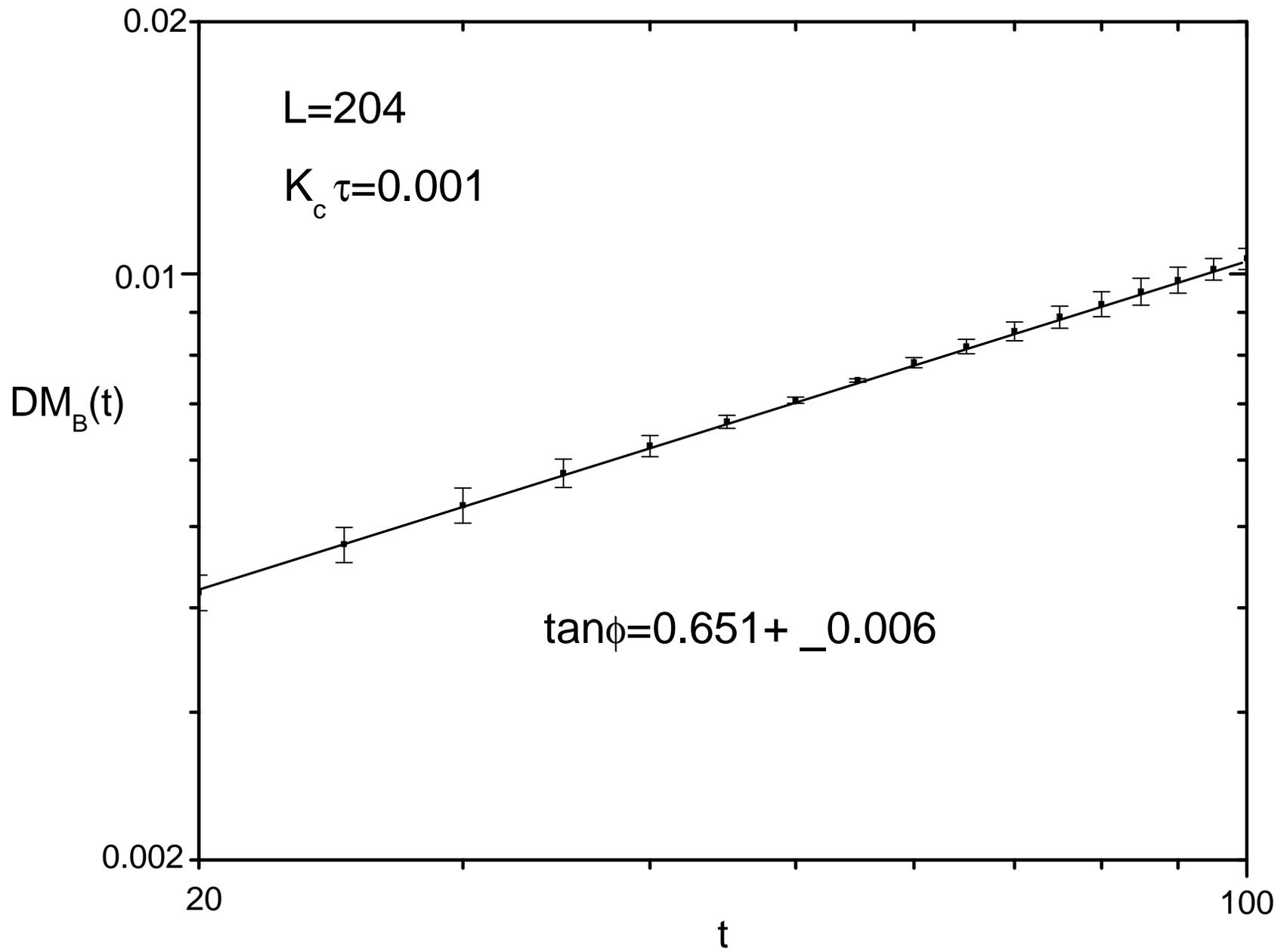
Figure 11. Logarithmic derivative of the Baxter's order parameter $M_{B}$.
Data were obtained for the lattice size $L=204$ using temperature variation $K_c \tau =0.001$. 
Five sets of $10000$ initially ordered samples ($m_0 = 1$) were used in order to 
obtain the error bars.

\end{document}